\documentclass[aip,jap,reprint,amsmath,amssymb]{revtex4-2}
\usepackage{graphicx}
\usepackage[colorlinks=true, allcolors=blue]{hyperref}
\usepackage[separate-uncertainty = true]{siunitx} 

\graphicspath{{Figures/}}
\DeclareSIUnit{\bar}{bar}
\newcommand{\SInobr}[2]{\SI[separate-uncertainty-units=single]{#1}{#2}}

\makeatletter
\def\@email#1#2{%
 \endgroup
 \patchcmd{\titleblock@produce}
  {\frontmatter@RRAPformat}
  {\frontmatter@RRAPformat{\produce@RRAP{*#1\href{mailto:#2}{#2}}}\frontmatter@RRAPformat}
  {}{}
}%
\makeatother

\begin{document}
\title{Minimization of micromotion for nanoparticles in a Paul trap}

\author{Jamie Morley}
\email{jamie.morley@uibk.ac.at}
\affiliation{Universit{\"a}t Innsbruck, Institut f{\"u}r Experimentalphysik, Technikerstra\ss e 25, 6020 Innsbruck, Austria}

\author{Jean Paul Louys Sansó}
\affiliation{Universit{\"a}t Innsbruck, Institut f{\"u}r Experimentalphysik, Technikerstra\ss e 25, 6020 Innsbruck, Austria}

\author{Dmitry Bykov}
\affiliation{Universit{\"a}t Innsbruck, Institut f{\"u}r Experimentalphysik, Technikerstra\ss e 25, 6020 Innsbruck, Austria}

\author{Simon Baier}
\affiliation{Universit{\"a}t Innsbruck, Institut f{\"u}r Experimentalphysik, Technikerstra\ss e 25, 6020 Innsbruck, Austria}

\author{Tracy Northup}
\affiliation{Universit{\"a}t Innsbruck, Institut f{\"u}r Experimentalphysik, Technikerstra\ss e 25, 6020 Innsbruck, Austria}

\date{\today}

\begin{abstract}
	When a charged particle in a Paul trap is displaced from the node of the AC trapping field, excess micromotion arises as an undesired effect. Excess micromotion heats the particle, limits the precision with which the particle can be localised, and acts as a decoherence channel in quantum mechanical experiments. However, thus far there is no standard procedure for micromotion compensation with mesoscopic particles.
	Here, we experimentally demonstrate three different methods for minimizing the micromotion of a nanoparticle in a linear Paul trap along three axes.
	The most precise method allows us to nullify the stray field to within \SI{2.9}{\volt\per\meter}, which is comparable to reported values in trapped-ion experiments.
\end{abstract}

\maketitle

Paul traps~\cite{paul1990electromagnetic} are a well-developed platform for trapping charged particles: electrons~\cite{matthiesen2021trapping}, single atomic ions and crystals thereof~\cite{leibfried2003quantum}, molecular ions~\cite{rellergert2013evidence,chou2017preparation, sinhal2020quantumnondemolition,wu2025infraredabsorptionspectroscopysingle}, highly charged ions~\cite{schmger2015coulomb}, nanoparticles~\cite{bonvin_hybrid_2024,bykov2019direct,tomassi2026accelerated,pontin2020ultranarrowlinewidth}, and microparticles~\cite{delord2020spincooling,message2025extremetemperature}.
A particle in a Paul trap exhibits both secular motion and micromotion~\cite{berkeland_minimization_1998};
excess micromotion arises when the stray field, which we define as the total DC field at the node of the AC trapping field, is nonzero.
Excess micromotion heats the particle, prevents high-precision localisation of the particle, and limits coherence in quantum mechanical experiments.
In order to nullify the stray field and thus the excess micromotion, DC voltages are applied to compensation electrodes.
Methods for detecting micromotion with trapped ions are based on a range of techniques, including resolved sidebands~\cite{keller_precise_2015,goham_resolved-sideband_2022,charles_doret_controlling_2012}, photon correlations~\cite{keller_precise_2015,berkeland_minimization_1998}, parametric excitation~\cite{ibaraki_detection_2011, keller_precise_2015,hogle2024precise,drewsen2004nondestructive,clark2021engineering} and RF power variation~\cite{saito_measurement_2021,saito_three-dimensional_2025}. The choice of technique depends on the context~\cite{keller_precise_2015}.

In recent years, Paul traps have emerged as a leading platform for levitodynamical experiments with nano- and microparticles~\cite{bonvin_hybrid_2024,bykov2019direct,delord2020spincooling,message2025extremetemperature,pontin2020ultranarrowlinewidth,tomassi2026accelerated}.
Excess micromotion introduces the same problems for
levitodynamical experiments as it does for trapped-ion experiments~\cite{dania2022position,stickler_quantum_2021,gonzalezballestero2021levitodynamics,gupta2026quantum,schut2026proposal}.
The micromotion detection methods developed for trapped ions are not in general directly transferable, due to the absence of a spectroscopic transition in mesoscopic particles.
Thus far, there is no standard procedure for minimizing micromotion within the levitodynamics community, although we note preliminary studies on the topic~\cite{bonvin_hybrid_2024,dania2021optical,dania2022position,penny2022lownoise,cavigelli2025toward}.

Here, we present and characterize three methods to minimize nanoparticle micromotion along three axes in linear Paul traps.
The tickling method, based on the response of the particle to a resonant modulation of the trapping potential, yields the highest compensation precision and is the most convenient to implement in the laboratory.
The radial stray field after using this method is \SI{0\pm2.9}{\volt\per\meter}.
We use this method to characterize different contributions to the stray field in our experimental system.
We then implement the tickling method in two additional systems, demonstrating the broad applicability of this method for micromotion compensation in a levitodynamics context.

\section{Background}\label{section:background}

\subsection{Micromotion in Paul traps}
Micromotion occurs in periodically driven systems such as Paul traps.
A Paul trap uses AC and DC electric fields to confine a charged particle dynamically.
The position components of the particle's center of mass $u_i(t)$ for $i \in \{x,~y,~z\}$ satisfy the Mathieu equation~\cite{berkeland_minimization_1998}.
The lowest-order solution, valid for Mathieu parameters $|a_i|\ll1, ~q_i^2\ll1$, is
\begin{multline}\label{eqn:eom_1d_solution}
	u_i(t)=\left[u_{0i}+u_{1i}\cos(\omega_{i}t+\Phi_i)\right]\left[1+\frac{q_i}{2}\cos(\Omega t)\right]
	,
\end{multline}
where $u_{0i}$ is a component of the equilibrium displacement from the AC field minimum $\vec{u}_0$,  $u_{1i}$ and $\omega_i$ are the amplitude and frequency of the secular motion, $\Phi_i$ is a phase dependent on the initial conditions, and $\Omega$ is the frequency of the AC field.
The secular frequencies can be approximated as $\omega_i \approx \frac{\Omega}{2}\sqrt{a_i + \frac{1}{2}q_i^2}$.

Equation~\eqref{eqn:eom_1d_solution} contains a secular motion term at $\omega_i$ with amplitude $u_{1i}$, two \textit{intrinsic} micromotion terms at $\Omega\pm\omega_i$ with amplitude $\frac{1}{4}u_{1i}q_i$, and an \textit{excess} micromotion term at $\Omega$ with amplitude $\frac{1}{2}u_{0i}q_i$.
The amplitude of the intrinsic micromotion can be reduced by cooling the secular mode or by lowering the $q_i$ parameter, e.g., by increasing $\Omega$. However, both $u_{1i}$ and $q_i$ are strictly nonzero, and therefore intrinsic micromotion is always present.

Excess micromotion is the subject of this work.
For a given set of trap and particle parameters  (fixed $q_i$), $u_{0i}$ needs to be minimized to minimize the amplitude of excess micromotion along the $i$-axis. This is done by counteracting the stray field component that causes ${u}_{0i} \neq 0$.

\begin{figure}[ht]
	\def\svgwidth{1.1\linewidth}
	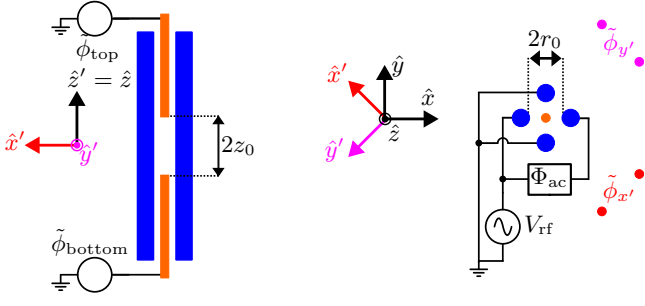
	\caption{Schematic of a linear Paul trap.
		One grounded electrode pair and one AC-driven electrode pair constitute the radial trapping electrodes.
		These electrodes are shown in blue, endcaps in orange, and compensation electrodes in pink and red.
		We displace a particle in three dimensions by varying the DC potentials $\tilde{\phi}_\text{top}, \tilde{\phi}_\text{bottom}, \tilde{\phi}_{x^\prime}, \tilde{\phi}_{y^\prime}$.
		Left: Side projection of the trap onto the $z$-$x^\prime$ plane.
		The distance between the surfaces of the top and bottom endcaps is $2z_0$.
		Right: Top projection of the trap onto the $x$-$y$ plane, including compensation electrodes.
		The distance between the surfaces of opposing radial trapping electrodes is $2r_0$.
		The amplitude of the applied AC potential between the pairs of radial trapping electrodes is $V_\text{rf}$. We include a phase shift $\Phi_\text{ac}$ between the potentials at the two AC-driven electrodes, assumed in this work to be negligible.
	}
	\label{fig:trap_schematic}
\end{figure}

Furthermore, if in the linear Paul trap shown in Fig.~\ref{fig:trap_schematic} we consider a nonzero phase shift $\Phi_{\text{ac}}\ll 1$ between the AC driven electrodes, an additional excess micromotion term $-\frac{q_i}{4}r_0\alpha \Phi_{\text{ac}}\sin(\Omega t)$ is present in $u_i(t)$, where $r_0$ is a characteristic length, and $\alpha$ is a geometric factor~\cite{berkeland_minimization_1998}.
Such a phase shift arises through experimental imperfections but is typically negligible, as we assume in this work.

\subsection{Excess micromotion minimization in trapped-ion systems}
Here we outline common methods for micromotion minimization in the trapped-ion community~\cite{berkeland_minimization_1998}.
These serve as a comparison point to the methods introduced in this work for micromotion minimization for nanoparticles.
For brevity, we label them Methods I--IV.

Method I is the resolved sideband method.
The amplitude of excess micromotion is determined by measuring the amplitude of the micromotion sideband relative to the carrier through spectroscopy of the atomic transition~\cite{goham_resolved-sideband_2022}.
Some reported stray field components~\footnote{Different authors choose to present the same information in different ways. When highlighting values reported in other works, we have kept the style from the source where appropriate. While the notation of $(0\pm \Delta E)$ conveys a clear message, we want to highlight that the measured quantity is the uncertainty $\Delta E$, with the assumption that the residual stray field is zero measured to a precision given by this uncertainty.} after compensation with this method are \SI{0\pm10}{\volt \per \meter}~\cite{charles_doret_controlling_2012} and \SI{0.6\pm0.6}{\volt \per \meter}~\cite{keller_precise_2015}.

Method II is the photon-correlation or cross-correlation method~\cite{berkeland_minimization_1998}.
The amplitude of excess micromotion and the phase with respect to the AC drive are extracted from the time delays between detected fluorescence and zero crossings of the drive.
Reference~\cite{keller_precise_2015} reported a stray field component after compensation with this method of \SI{0.09\pm0.09}{\volt \per \meter}.

Method III is the parametric resonance or tickling method~\cite{ibaraki_detection_2011, keller_precise_2015,hogle2024precise,drewsen2004nondestructive,clark2021engineering}. The amplitude of excess micromotion is extracted from the fluorescence response to a modulation of the drive at $\omega_m=2\omega_i/n$ for $n\in\mathbb{N}$~\cite{ibaraki_detection_2011}.
Method III allows for micromotion compensation along three axes with just one laser.
Some reported stray field components after compensation with this method are \SI{0\pm6}{\volt \per \meter}~\cite{tanaka_micromotion_2012} and \SI{0\pm0.3}{\volt \per \meter}~\cite{keller_precise_2015}.

Other methods to minimize micromotion in trapped-ion systems exist, such as RF power variation~\cite{saito_measurement_2021, saito_three-dimensional_2025}, using ultracold clouds of neutral atoms~\cite{harter_minimization_2013}, using high finesse cavities~\cite{chuah_detection_2013}, or analyzing the ion's trajectory~\cite{gloger_ion-trajectory_2015}.
Combinations of the above methods are also used in situations in which laser access is restricted~\cite{tanaka_micromotion_2012}.
Micromotion minimization can be sped up by using machine learning~\cite{liu_minimization_2021}.

\section{Excess micromotion minimization for nanoparticles}

In this section we introduce three methods that were used to detect and subsequently compensate for micromotion in our system.
The methods described here are based on detecting the micromotion amplitude directly (Method 1), measuring the phase of the micromotion (Method 2) and mapping the trapping field via tickling (Method 3), and are deliberately indexed such as to show similarity to the trapped-ion Methods I--III.

\subsection{Experimental setup}
The linear Paul trap used in this work is shown schematically in Fig.~\ref{fig:trap_schematic}. The distance between the surfaces of opposing AC-driven electrodes is $2r_0=\SI{1.8}{\milli \meter}$, and the distance between the surfaces of the endcaps is $2z_0=\SI{3.2}{\milli\meter}$.
The trap axis ($z$-axis) is vertical.
We trapped a silica sphere with a nominal diameter of \SI{300}{\nano \meter}, at a pressure of around $\SI{e-2}{\milli\bar}$, with a drive frequency $\Omega/2\pi=\SI{7}{\kilo\hertz}$ and with a peak-to-peak amplitude $2V_\text{rf}=\SI{450}{\volt}$.
The mass of the particle was not determined.
A voltage of \SI{70}{\volt} was applied to both endcaps before any deliberate displacement of the particle; this value was high enough to render the axial displacement due to gravity negligible.
These parameters led to eigenfrequencies $(\omega_z, \omega_x, \omega_y)/2\pi = (610, 798, 886)~\si{\hertz}$.
All data were taken with the same particle, using these trap parameters, and without center-of-mass cooling, unless explicitly stated.

The particle was displaced in the radial plane by applying DC voltages $(\tilde{\phi}_{x^\prime},\tilde{\phi}_{y^\prime})$ to two pairs of compensation electrodes (see Fig.~\ref{fig:trap_schematic}).
These potentials had a negligible effect on the trap stiffness in any direction.
The particle was displaced along the trap axis by changing the relative endcap potential $\tilde{\phi}_{z^\prime}=\tilde{\phi}_{\text{bottom}}-\tilde{\phi}_{\text{top}}$, where $\tilde{\phi}_\text{bottom}$ and $\tilde{\phi}_\text{top}$ were the potentials applied to the bottom and top endcaps (see Appendix~\ref{appendix:axial_displacement}).
The DC fields that resulted from applying these potentials were close to orthogonal (see Appendix~\ref{appendix:transformation}).
Experimentally, the values $(\phi_{x^\prime}, \phi_{y^\prime}, \phi_{\text{bottom}}, \phi_{\text{top}})$ were the setpoints of the DC source (EHS 8220n, \textit{iseg Spezialelektronik GmbH}); these voltages were subsequently stepped down by a factor of approximately $1/6$ at a low-pass filtering stage before the electrodes, such that $\tilde{\phi}_k \approx \phi_k/6$ for every setpoint $\phi_k$.

\begin{figure}[ht]
	\includegraphics[width=1\linewidth]{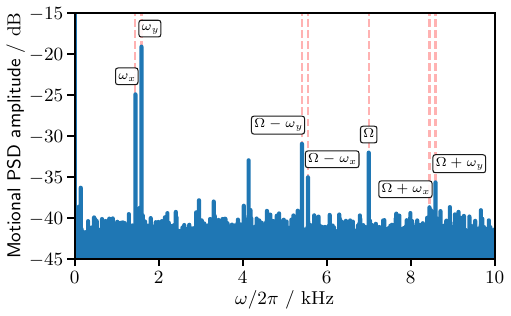}
	\caption{
		A typical PSD before micromotion compensation, obtained from the detection system most sensitive to the $x^\prime$ projection of the particle's motion.
		Motional peaks are visible at $\omega_{i}, \Omega, \Omega\pm \omega_{i}, i \in (x,y)$.
		The peak at $\Omega + \omega_x$ is barely distinguishable from the noise floor, and the peak at \SI{4141}{\hertz} is electronic noise.
	}
	\label{fig:motional_spectrum}
\end{figure}

The particle's motion was detected optically, via three separate confocal detection systems that collect light scattered from the particle~\cite{dania2024ultrahigh}.
Each system was optimized to be most sensitive to motion along only one of the $x^\prime, y^\prime$ and $z^\prime$ axes, which form a basis rotated \SI{45}{\degree} about the trap axis relative to the eigenbasis (see Fig.~\ref{fig:trap_schematic}).
The detection systems were not calibrated, so measured motional amplitudes are given in decibels relative to a value of \SI{1}{\volt^2 \per\hertz}.
A typical power spectral density (PSD) of the detected signal before micromotion compensation is shown in Fig.~\ref{fig:motional_spectrum}.

\begin{figure}[ht]
	\includegraphics[width=1\linewidth]{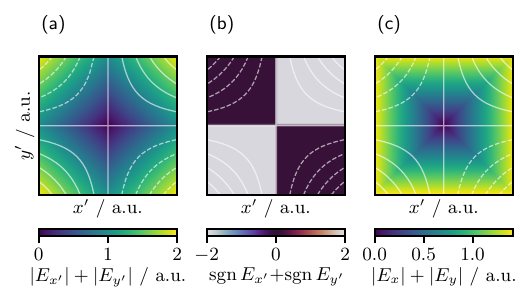}
	\caption{Simulations of an ideal quadrupole field $\vec{E}$ generated by the radial trapping electrodes in Fig.~\ref{fig:trap_schematic}.
		The simulations are centered at the node for a given time; displacement of a trapped particle from the node causes excess micromotion.
		Equipotential lines are shown in white.
		(a) Sum of the absolute values of the projections $E_{x^\prime}\equiv\vec{E}\cdot \hat{x}^\prime$ and $E_{y^\prime}\equiv\vec{E}\cdot \hat{y}^\prime$.
		(b) Sum of the signs of the projections $E_{x^\prime}$ and $E_{y^\prime}$.
		(c) Sum of the absolute values of the projections $E_{x}\equiv\vec{E}\cdot \hat{x}$ and $E_{y}\equiv\vec{E}\cdot \hat{y}$.
	}
	\label{fig:theory_maps}
\end{figure}

\subsection{Method 1: Direct amplitude detection}
In Method 1, we extract the amplitude $A_{i^\prime \in \{x^\prime, y^\prime, z^\prime \}}$ of the excess micromotion peak at $\Omega$ from the measured PSD; see, for example, Fig.~\ref{fig:motional_spectrum}, in which $A_{x^\prime}= \SI{-32}{\decibel}$.
Due to the quadrupole symmetry in the trap's radial plane, $A_{x^\prime}$ is minimized at $u_{0y^\prime} \equiv \vec{u}_0\cdot \hat{y}^\prime=0$; analogously, $A_{y^\prime}$ is minimized at $u_{0x^\prime} \equiv \vec{u}_0\cdot \hat{x}^\prime=0$.
The radial micromotion minimum is extracted from measurements of $A_{x^\prime}$ and $A_{y^\prime}$ for different $(\phi_{x^\prime}, \phi_{y^\prime})$ pairs, keeping $\phi_{z^\prime}$ constant.
The profile of $A_{x^\prime}+A_{y^\prime}$ is expected to resemble that of $\left|E_{x^\prime}\right|+\left|E_{y^\prime}\right|$, where $E_{i^\prime}$ is the projection of the AC field along the $i^\prime$ axis. The profile of $\left|E_{x^\prime}\right|+\left|E_{y^\prime}\right|$ for an ideal quadrupole field is shown in Fig.~\ref{fig:theory_maps}(a).
The axial micromotion minimum (at $u_{0z^\prime} \equiv \vec{u}_0\cdot \hat{z}^\prime=0$) is extracted from a measurement of $A_{z^\prime}$ for different $\phi_{z^\prime}$, keeping $(\phi_{x^\prime}, \phi_{y^\prime})$ constant.

For micromotion minimization, Method 1 is intuitive and easy to implement along one axis.
However, minimization along three axes requires three detection systems, each sensitive to motion along one axis, for the best results.
Additionally, drifts in the detection sensitivity can skew the results obtained via Method 1.

\subsection{Method 2: Direct phase detection}
In Method 2, we extract the phase $\Phi_{i^\prime \in \{x^\prime, y^\prime, z^\prime \}}$ of the micromotion relative to the AC drive.
Due to the quadrupole symmetry in the trap's radial plane, $\Phi_{x^\prime}$ undergoes a $\pi$ shift at $u_{0y^\prime} =0$; analogously, $\Phi_{y^\prime}$ undergoes a $\pi$ shift at $u_{0x^\prime} =0$.
The radial micromotion minimum is extracted from measurements of $\Phi_{x^\prime}$  and $\Phi_{y^\prime}$ for different $(\phi_{x^\prime}, \phi_{y^\prime})$ pairs, keeping $\phi_{z^\prime}$ constant.
The profile of $\left(\Phi_{x^\prime}+\Phi_{y^\prime}\right)\pmod{2\pi}$ is expected to resemble that of $\text{sgn}~E_{x^\prime}+\text{sgn}~E_{y^\prime}$. The profile of $\text{sgn}~E_{x^\prime}+\text{sgn}~E_{y^\prime}$ for an ideal quadrupole field is shown in Fig.~\ref{fig:theory_maps}(b).
The axial micromotion minimum (at $u_{0z^\prime} =0$) is extracted from a measurement of $\Phi_{z^\prime}$ for different $\phi_{z^\prime}$, keeping $(\phi_{x^\prime}, \phi_{y^\prime})$ constant.

For micromotion minimization, Method 2 does not require the detection to be calibrated for any drifts.
However, as for Method 1, this method requires three detection systems, each sensitive to motion along one axis, for best results.

\begin{figure*}
	\includegraphics[width=1\linewidth]{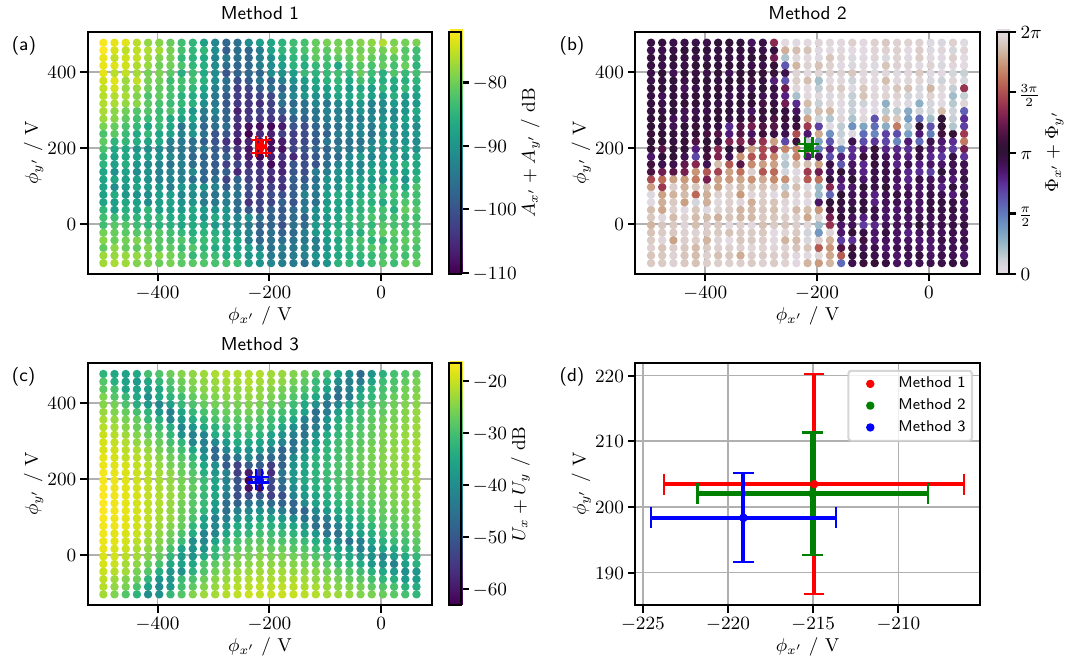}
	\caption{Data from applying Methods 1--3 to find the micromotion minimum in the radial plane.
		The setpoints $\phi_{x^\prime}$ and $\phi_{y^\prime}$ are shown on the horizontal and vertical axes, and their ranges in (a--c) correspond to scanning an area of approximately $(0.014 r_0)^2$.
		(a) Method 1: sum of the excess micromotion amplitudes $A_{x^\prime}$ and $A_{y^\prime}$. The extracted micromotion minimum is shown in red.
		(b) Method 2: sum of the excess micromotion phases $\Phi_{x^\prime}$ and $\Phi_{y^\prime}$. The extracted micromotion minimum is shown in green.
		(c) Method 3: sum of the secular-motion amplitudes $U_{x}$ and $U_{y}$ during simultaneous tickling of the $x$ and $y$ modes. The extracted micromotion minimum is shown in blue.
		(d) Radial excess micromotion minima found via Methods 1--3.
	}
	\label{fig:method_123_2d}
\end{figure*}

\subsection{Method 3: Resonant tickling}
In Method 3, we modulate the amplitude of the AC trap drive.
Resonant excitation of mode $i \in \{x, y, z\}$ occurs when the modulation frequency $\Omega_{\text{tickle},i}$ is equal to $\Omega \pm \omega_i$ (see Appendix~\ref{appendix:tickling_frequencies}) and there is a nonzero projection of the AC field along $\hat{i}$.
The amplitude of the AC potential is modulated according to
\begin{equation}\label{eqn:bichromatic_tickling}
	V_\text{rf}(t) = V_\text{rf}(0)\left(1+\sum_i h_i \sin(\Omega_{\text{tickle},i}t)\right),
\end{equation}
where $h_i \ll 1$ is the modulation depth at $\Omega_{\text{tickle},i}$. The potential applied at the AC-driven electrodes is then $V_\text{rf}(t)\sin(\Omega t)$. Equation~\ref{eqn:bichromatic_tickling} shows that, while not strictly necessary, Method 3 allows for simultaneous tickling of multiple modes.

From the measured PSD, we extract the amplitude $U_{i}$ of the secular-motion peak at $\omega_i$; see, for example, Fig.~\ref{fig:motional_spectrum}, where $U_{x}= \SI{-25}{\decibel}$.
Due to the quadrupole symmetry in the trap's radial plane, $U_{x}$ is minimized during tickling of the $x$-mode at $u_{0x} =0$; analogously, $U_{y}$ is minimized during tickling of the $y$-mode at $u_{0y} =0$.
The radial micromotion minimum is extracted from measurements of $U_{x}$ and $U_{y}$ for different $(\phi_{x^\prime}, \phi_{y^\prime})$ pairs while tickling the $x$ and $y$ modes simultaneously and keeping $\phi_{z^\prime}$ constant.
The profile of $U_{x}+U_{y}$ is expected to resemble that of $\left|E_{x}\right|+\left|E_{y}\right|$, where $E_{i}$ is the projection of the AC field along the $i$ axis. The profile of $\left|E_{x}\right|+\left|E_{y}\right|$ for an ideal quadrupole field is shown in Fig.~\ref{fig:theory_maps}(c).
The axial micromotion minimum (at $u_{0z^\prime} =0$) is extracted from a measurement of $U_{z}$ for different $\phi_{z^\prime}$ while tickling the $z$ mode and keeping $(\phi_{x^\prime}, \phi_{y^\prime})$ constant.

Method 3 has three advantages over Methods 1 and 2.
First, its precision is limited by the thermal noise rather than our detection noise floor.
Second, optimal compensation along three axes can be achieved with a single detection system.
Third, Method 3 relies on the $\omega_i$ peaks instead of the $\Omega$ peak, which separates the micromotion amplitude information for each axis.

\section{Experimental demonstration of minimization methods}

\subsection{Finding the radial micromotion minimum}
Figure~\ref{fig:method_123_2d} is the main result of this work. It shows data from applying Methods 1--3 to find the micromotion minimum in the radial plane.

Figure~\ref{fig:method_123_2d}(a) shows the sum $A_{x^\prime}+A_{y^\prime}$ as a function of $\phi_{x^\prime}$ and $\phi_{y^\prime}$.
The profile we obtained is consistent with that of Fig.~\ref{fig:theory_maps}(a).
In the dark region, the measured micromotion amplitudes are at the detection noise floor.
However, sub-noise precision in the micromotion minimum can be obtained.
Here, we extracted the minimum $(\phi_{x^\prime},\phi_{y^\prime})=(\SInobr{-215(9)}{}, \SInobr{203(20)}{})~\si{\volt}$.
Appendix~\ref{appendix:find_optimum} provides more information on sub-noise precision and describes how the minima were obtained for this plot and others.

Figure \ref{fig:method_123_2d}(b) shows the sum $\Phi_{x^\prime} + \Phi_{y^\prime}$ as a function of $\phi_{x^\prime}$ and $\phi_{y^\prime}$.
The profile we obtained is consistent with that of Fig.~\ref{fig:theory_maps}(b).
Near the phase boundaries, the micromotion can no longer be resolved, so the measured phase becomes noisy.
Again, sub-noise precision in the micromotion minimum can be obtained (see Appendix~\ref{appendix:find_optimum}).
Here, we extracted the minimum $(\phi_{x^\prime},\phi_{y^\prime})=(\SInobr{-215(7)}{}, \SInobr{202(9)}{})~\si{\volt}$.

Figure~\ref{fig:method_123_2d}(c) shows the sum $U_{x} + U_{y}$ as a function of $\phi_{x^\prime}$ and $\phi_{y^\prime}$ with simultaneous tickling at $\Omega-\omega_x$ and $\Omega-\omega_y$. The tickling depths were $h_x,~h_y\approx \SI{1}{\percent}$.
The profile we obtained is consistent with that of Fig.~\ref{fig:theory_maps}(c).
In the dark region, the excitation of both secular modes is minimized.
The data in Fig.~\ref{fig:method_123_2d}(c) appear less noisy than for the other methods, due to the reasons we discussed above in presenting Method 3.
Here, we extracted the minimum $(\phi_{x^\prime},\phi_{y^\prime})=(\SInobr{-219(5)}{}, \SInobr{198(7)}{})~\si{\volt}$.

The optimal $\phi_{x^\prime},~\phi_{y^\prime}$ pairs extracted from the three methods are shown together in Fig.~\ref{fig:method_123_2d}(d).
Details of how these voltages are found from the data are given in Appendix~\ref{appendix:find_optimum}.
The values obtained from all methods are in agreement.
Figure~\ref{fig:method_123_2d}(d) also shows the precision of the three methods; Method~3 is seen to be the most precise, followed by Method~2 and then Method~1.

To allow for direct comparison with other systems, three metrics for the micromotion minimization precision are shown in Table~\ref{tab:precision_metrics} (see Appendix~\ref{appendix:precision_metrics}).
The smallest uncertainty of the residual stray field is found for Method 3 at \SI{2.9}{\volt\per\meter}, which is comparable to values found for trapped ions~\cite{charles_doret_controlling_2012,tanaka_micromotion_2012,saito_measurement_2021}.
For this reason, Method~3 has since been used in two other Paul traps in our laboratory, for nanoparticles of different sizes and materials (see Appendix~\ref{appendix:diamond_trap}).

\begin{table}[ht]
	\centering
	\caption{Precision metrics for the radial data presented in Fig.~\ref{fig:method_123_2d}. For each metric, we report the largest value across the two dimensions $x^\prime$ and $y^\prime$.
	}
	\label{tab:precision_metrics}
	\begin{tabular}{|c|c|c|c|}
		\hline
		Method & Localization                   & Residual excess               & Residual stray field                 \\

		       & precision / \si{\micro \meter} & micromotion                   & uncertainty / \si{\volt \per \meter} \\
		       &                                & amplitude / \si{\nano \meter} &                                      \\
		\hline
		1      & 0.6                            & 108                           & 7.2                                  \\
		\hline
		2      & 0.58                           & 104                           & 4                                    \\
		\hline
		3      & 0.57                           & 103                           & 2.9                                  \\
		\hline
	\end{tabular}
\end{table}

\begin{figure}[ht]
	\includegraphics[width=1\linewidth]{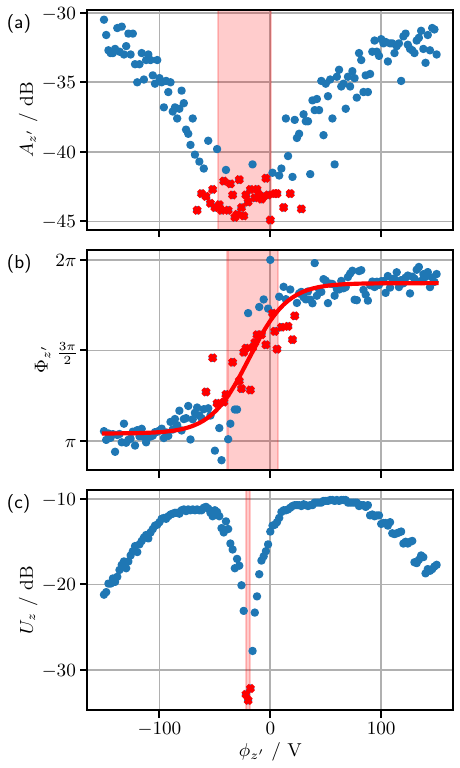}
	\caption{Data from applying Methods 1--3 to find the micromotion minimum along the $z$ axis.
		The setpoint $\phi_{z^\prime}$ is shown on the horizontal axes, and its range corresponds to approximately \SI{180}{\micro\meter}.
		The extracted micromotion minima are shown as red-shaded regions.
		(a) Method 1: the excess micromotion amplitude $A_{z^\prime}$.
		(b) Method 2: the excess micromotion phase $\Phi_{z^\prime}$.
		(c) Method 3: the secular-motion amplitude $U_{z}$ during tickling of the $z$ mode.
	}
	\label{fig:z_compensation}
\end{figure}

\subsection{Finding the axial micromotion minimum}
We now turn to finding the micromotion minimum along the trap axis.
As the amplitude of axial micromotion was much smaller than the amplitude of radial micromotion, we increased the peak-to-peak amplitude of the drive to \SI{960}{\volt}, leading to eigenfrequencies $(\omega_z, \omega_x, \omega_y)/2\pi = (622, 2112, 2368)~\si{\hertz}$.
For each method, we do not extract the minimum from a fit, since we make no assumptions about the functional form of each profile in Fig.~\ref{fig:z_compensation}.
Instead, we identify a set of data points that corresponds to the minimum, shown in red.

Figure~\ref{fig:z_compensation}(a) shows $A_{z^\prime}$ as a function of $\phi_{z^\prime}$.
We see a v-shaped profile with a broad minimum.
All measured amplitudes were within the linear response range of our detection system.
The red points correspond to amplitudes within \SI{3}{\deci\bel} fluctuations of the minimum value.
From the mean and standard deviation of these points, we extracted the minimum $\phi_{z^\prime}=\SI{-23(24)}{\volt}$.

Figure~\ref{fig:z_compensation}(b) shows $\Phi_{z^\prime}$ as a function of $\phi_{z^\prime}$.
We see two plateaus separated by a region where the measured phase is noisy.
In order to extract the phase jump~$\Delta$ between the plateaus, we fit a sigmoid function; the obtained value $\Delta  = \SI{0.83(3)}{\pi}$ is lower than the expected value of \SI{}{\pi}.
The red points are all those at least $0.2\Delta$ above the lower plateau and at least $0.2\Delta$ below the upper plateau.
From the mean and standard deviation of these points, we extracted the minimum $\phi_{z^\prime}=\SI{-16(23)}{\volt}$.

Figure~\ref{fig:z_compensation}(c) shows $U_{z}$ as a function of $\phi_{z^\prime}$.
We see a profile like the Aries glyph, with a sharp minimum.
The drop in $U_{z}$ for values $\left|\phi_{z^\prime}\right| \gtrsim \SI{80}{\volt}$ is due to both a decreased sensitivity and a nonlinear response in the detection system.
The red points correspond to amplitudes within \SI{3}{\deci\bel} fluctuations of the minimum value.
From the mean of these points and the scan resolution, we extracted the minimum $\phi_{z^\prime}=\SI{-20(2)}{\volt}$.

The minima obtained from all methods are in agreement.
We highlight that individual  and vertical  slices of the datasets added together in Figs.~\ref{fig:method_123_2d}(a--c) (see Appendix~\ref{appendix:find_optimum}) have the same structure as the profiles (constant $\phi_{x^\prime}, \phi_{y^\prime}$) shown in Figs.~\ref{fig:z_compensation}(a--c), which is expected.
The detected amplitudes of excess micromotion, $A_{i^\prime}$, reach the noise floor in Figs.~\ref{fig:method_123_2d}(a) and \ref{fig:z_compensation}(a).
This places an upper bound $\Phi_\text{ac}\leq \frac{2u_{0i}}{r_0 \alpha}\approx \SInobr{1e-4}{}$ for typical values of $\alpha$~\cite{berkeland_minimization_1998}, which validates the assumption that $\Phi_\text{ac}$ is negligible.
As in Fig.~\ref{fig:method_123_2d}, Method 3 is seen in Fig.~\ref{fig:z_compensation} to be the most precise.

\begin{figure}[hb]
	\includegraphics[width=1\linewidth]{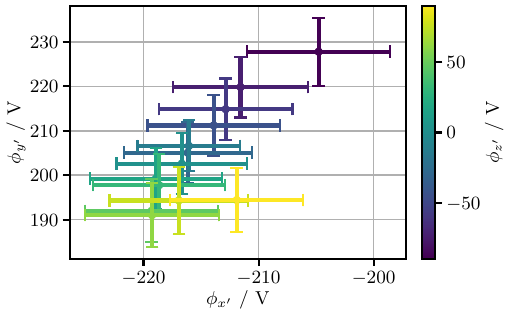}
	\caption{Radial minimum obtained via Method 3 for different $\phi_{z^\prime}$.
		The setpoints $\phi_{x^\prime}$ and $\phi_{y^\prime}$ are shown on the horizontal and vertical axes, and the setpoint $\phi_{z^\prime}$ is shown on the colorbar.
		The range of $\phi_{z^\prime}$ corresponds to approximately \SI{90}{\micro\meter}.
	}
	\label{fig:method_3_over_z}
\end{figure}

\section{Characterization of the stray field}
In this section, we characterize contributions to the stray field in our system by studying changes in the radial micromotion minimum.
Only data obtained via Method 3 are presented, due to its higher precision and simpler practical implementation.

Figure~\ref{fig:method_3_over_z} shows the radial minimum for different $\phi_{z^\prime}$.
For $\left|\phi_{z^\prime}\right| < \SI{50}{\volt}$, we cannot identify any change in the radial component of the stray field.
For $\left|\phi_{z^\prime}\right| > \SI{50}{\volt}$, there is a $2\sigma$ variation in the minimum. The two endcaps of our trap are radially displaced from one another by approximately one endcap radius (\SI{0.25}{\milli\metre}), which we suspect caused the observed variation.

\begin{figure*}
	\includegraphics[width=1\linewidth]{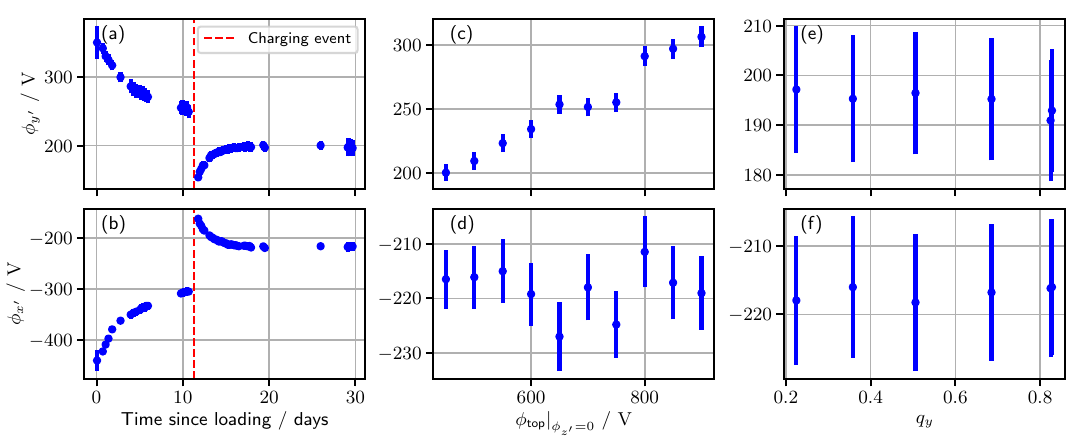}
	\caption{The $\phi_{y^\prime}$ and $\phi_{x^\prime}$
		components of the radial minimum, obtained via Method 3, as a function of (a,~b) the time since the particle was loaded;
		(c,~d)  $\phi_\text{top}$, for $\phi_{z^\prime}=0$;
		(e,~f) the Mathieu parameter $q_y$.
		Eleven days after loading the particle, there was a charging event, indicated in red in (a) and (b).
	}
	\label{fig:radial_comp_over_param_space}
\end{figure*}

Figures~\ref{fig:radial_comp_over_param_space}(a) and \ref{fig:radial_comp_over_param_space}(b) show the $y^\prime$ and $x^\prime$ components of the radial minimum as a function of the time since the particle was loaded.
Eleven days after loading the particle, we connected a pressure gauge to the vacuum chamber; we attribute the jumps in the figures to a change in the local charge environment caused by the gauge~\cite{dania2021optical}. Before and after the charging event, the compensation voltages required to nullify the stray field are seen to relax towards stable values.
We interpret this as a slow stabilization of the local charge environment~\footnote{The data in Figs.~\ref{fig:radial_comp_over_param_space}(a) and \ref{fig:radial_comp_over_param_space}(b) suggest the stray field primarily changed along the $y$-axis. This directionality, as well as the relaxation mechanism of the stray field, are beyond the scope of this work.}.
Around one week after the charging event, the stray field reached a steady state.
All of the data in this work, with the exception of those shown in Figs.~\ref{fig:radial_comp_over_param_space}(a) and \ref{fig:radial_comp_over_param_space}(b), were acquired only after reaching the steady state.

Figures~\ref{fig:radial_comp_over_param_space}(c) and \ref{fig:radial_comp_over_param_space}(d) show the $y^\prime$ and $x^\prime$ components of the radial minimum as a function of $\phi_\text{top}$, for $\phi_{z^\prime}=0$.
The value of $\phi_{x^\prime}$ remained constant, whereas the value of $\phi_{y^\prime}$ increased significantly as $\phi_\text{top}$ was increased. Based on this observation, we hypothesize that the endcap axis is radially displaced from the AC node by $\Delta x^\prime \approx 0$, $\Delta y^\prime>0$.

Figures~\ref{fig:radial_comp_over_param_space}(e) and \ref{fig:radial_comp_over_param_space}(f) show the $y^\prime$ and $x^\prime$ components of the radial minimum as a function of $q_y$, which had the largest absolute value of the Mathieu parameters $\{q_x, q_y, q_z\}$.
We increased $q_y$ by increasing the amplitude of the AC drive; the range of $0.2 < q_y < 0.9$ corresponds to a peak-to-peak AC drive amplitude in the range (360 to 870)~\si{\volt}.
The values of both $\phi_{x^\prime}$ and $\phi_{y^\prime}$ remained constant, as expected, since changing $q_i$ should not change the stray field.

\section{Conclusion and Outlook}

Motivated by the lack of a standard procedure, we have introduced and demonstrated three methods for the detection and subsequent compensation of micromotion for nanoparticles in a Paul trap.
The methods are based on direct amplitude detection, direct phase detection, and resonant tickling.
We have shown that all three methods are in agreement and allow for micromotion minimization along three axes.

The tickling method was the most precise and is also the most convenient, since it requires only one detection system.
This method can be used in any Paul trap in which the thermal amplitudes of the trapped particle's secular modes can be resolved.
Via this method, we have characterized the stray field in our system and nullified its radial component to within \SI{2.9}{\volt \per \meter}, which is comparable to values reported for trapped-ion systems.

We anticipate that a higher precision in each method presented here can be achieved by increasing the detection efficiency, fitting each profile, cooling the center-of-mass motion, and reducing the pressure in the vacuum chamber.
Furthermore, we have not examined how to use these methods to minimize micromotion efficiently.
The methods presented here provide a basis for developing a standard procedure for micromotion minimization within the levitodynamics community.

The datasets and analysis code are available at https://doi.org/10.5281/zenodo.21620140.

\begin{acknowledgments}
	This research was funded in part by the Austrian Science Fund (FWF) [10.55776/COE1, 10.55776/I5540, 10.55776/ESP258]. For Open Access purposes, the authors have applied a CC BY public copyright license to any author accepted manuscript version arising from this submission.
\end{acknowledgments}

\bibliographystyle{aipnum4-2}
\bibliography{bibliography}

\begin{figure*}
	\includegraphics[width=1\linewidth]{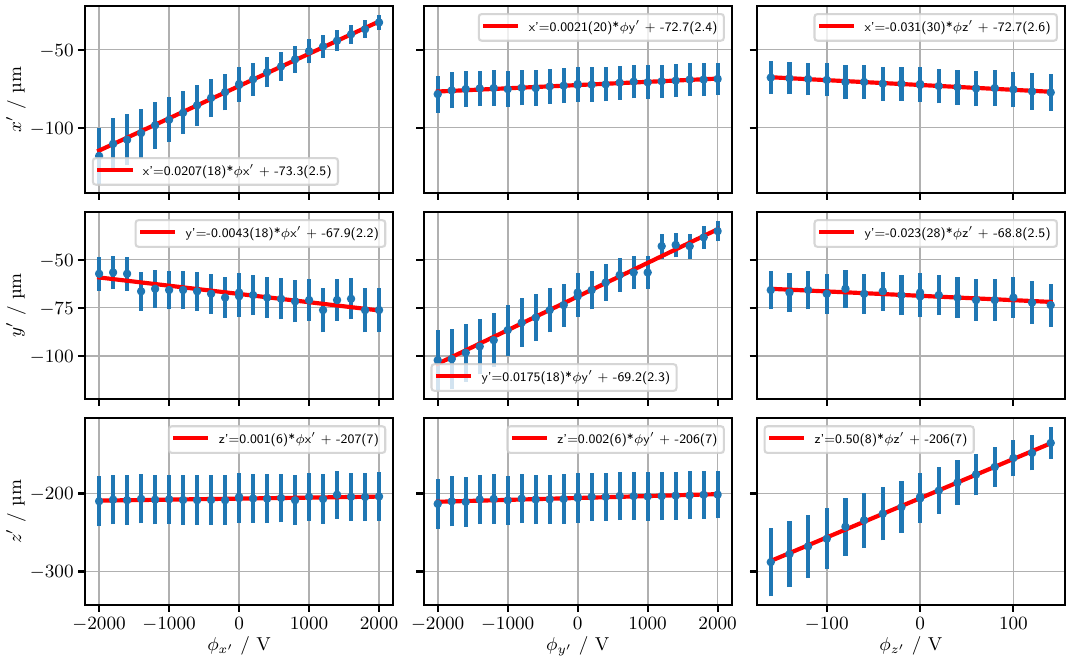}
	\caption{Particle's spatial coordinates $x^\prime$ (top row), $y^\prime$ (middle row), $z^\prime$ (bottom row) as a function of each setpoint $\phi_{x^\prime}$ (left column), $\phi_{y^\prime}$ (middle column), $\phi_{z^\prime}$ (right column).
		Data are shown in blue, and linear fits are shown in red.
		The origin $\left( x^\prime, y^\prime, z^\prime\right)=(0,0,0)$ was arbitrarily chosen, and in each column the other setpoints were held at $\SI{0}{\volt}$.
	}
	\label{fig:voltage_position_matrix}
\end{figure*}

\appendix
\section{Axial displacement at constant stiffness}\label{appendix:axial_displacement}
In order to displace the particle along the trap axis, we need to apply different potentials to each endcap.
However, arbitrary endcap potentials will in general change the axial stiffness, that is, the second derivative of the axial trapping potential.
For our system, we saw that a constant geometric mean $\phi_\text{end}\equiv\sqrt{\phi_\text{top} \phi_\text{bottom}}$ maintains a constant stiffness.
The value of $\omega_z$ was seen to stay constant over a displacement $\phi_{z^\prime}$ in the range $(-0.44\phi_\text{end}$, $0.44\phi_\text{end})$ at a pressure of  $\SI{e-2}{\milli\bar}$, where the linewidth was approximately \SI{10}{\hertz}.

Note that when the trap axis is vertical, as in our case, the value $\phi_\text{end}$ should be chosen such that the axial confinement is not significantly altered by gravity.
We do this by increasing $\phi_\text{end}$ for $\phi_{z^\prime}=0$ until the particle is no longer vertically displaced.

\section{Voltage-to-position transformation}\label{appendix:transformation}
Figure~\ref{fig:voltage_position_matrix} shows the particle's position $\vec{r}=(x^\prime,y^\prime,z^\prime)$ as a function of $\vec{\phi}=(\phi_{x^\prime},\phi_{y^\prime},\phi_{z^\prime})$, which we assume to satisfy $\vec{r}=\mathbf{A} \vec{\phi}+\vec{r}_0$.
The position was extracted from two orthogonal cameras, and the origin $\vec{r}=\vec{0}$ was arbitrarily chosen.
Error bars in the position arise from uncertainties in the camera calibrations.

From the data shown in Fig.~\ref{fig:voltage_position_matrix}, we extract $\vec{r}_0=(\SInobr{72.9(2.5)}{},\SInobr{68.6(2.3)}{},\SInobr{206(7)}{})~\si{\micro \metre}$ and reconstruct the matrix
$$\mathbf{A}=
	\left(\begin{matrix}\SInobr{20.7(1.8)}{} & \SInobr{2.1(2.0)}{} & \SInobr{31(30)}{} \\\SInobr{4.3(1.8)}{} & \SInobr{17.5(1.8)}{} & \SInobr{23(28)}{}\\ \SInobr{1(6)}{} & \SInobr{2(6)}{} & \SI{5.0(8)e+02}{}
	\end{matrix}\right)~\si{\nano \meter \per \volt}$$
via $A_{ij} \equiv \frac{\partial r_i}{\partial \phi_j}$.
The particle is significantly more sensitive to $\phi_{z^\prime}$ than to other components of $\vec{\phi}$, as expected from finite element simulations.
This transformation is dependent on trap parameters (see, for example, Eqn.~16 of~\cite{berkeland_minimization_1998}), and $\mathbf{A}$ was obtained under the default parameters used in this work.

\section{Tickling frequencies}\label{appendix:tickling_frequencies}
Here we discuss the excitation of the particle due to different tickling frequencies $\Omega_\text{tickle}$.
By considering the force from the AC field as a perturbation to harmonic particle dynamics, it can be shown that Eqn.~\ref{eqn:bichromatic_tickling} gives rise to leading-order force terms at frequencies $\Omega \pm \Omega_\text{tickle}$ in the one-dimensional case.
Hence a modulation at $\Omega_{\text{tickle}}=\Omega - \omega_i$ leads to a force term at $\omega_i$. The resulting excitation is stronger than that from the force terms at $\Omega \pm \omega_i$ arising from $\Omega_{\text{tickle}}=\omega_i$.
For this reason, we chose $\Omega_{\text{tickle}}=\Omega - \omega_i$.
We note, however, that as in trapped-ion experiments, excitation was experimentally observed in our system for $\Omega_{\text{tickle}}=2\omega_i/n$, $n\in \{1,2,3\}$, and we anticipate that $\Omega_{\text{tickle}}=\Omega - 2\omega_i/n$ could lead to similar behavior.

\section{Extraction of the minimum in the radial plane}\label{appendix:find_optimum}
\begin{figure}[h!]
	\includegraphics[width=1\linewidth]{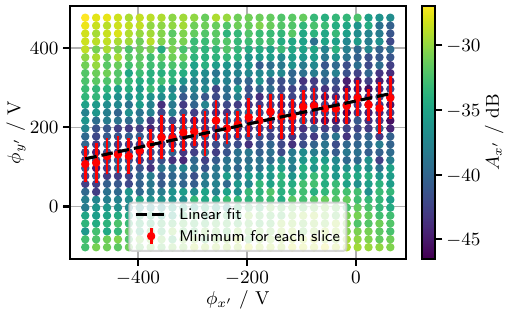}
	\caption{
		Data from applying Method 1 to find the micromotion minimum along the $y^\prime$ axis.
		The setpoints $\phi_{x^\prime}$ and $\phi_{y^\prime}$ are shown on the horizontal and vertical axes, and the excess micromotion amplitude $A_{x^\prime}$ is shown in the color bar.
	}
	\label{fig:radial_method1_extraction}
\end{figure}

The plots shown in Fig.~\ref{fig:method_123_2d} are presented as sums of two different two-dimensional profiles.
For example, Fig.~\ref{fig:method_123_2d}(a) shows $A_{x^\prime}(\phi_{x^\prime},\phi_{y^\prime})+ A_{y^\prime}(\phi_{x^\prime},\phi_{y^\prime})$.
One of these two-dimensional profiles, $A_{x^\prime}(\phi_{x^\prime},\phi_{y^\prime})$, is shown in Fig.~\ref{fig:radial_method1_extraction}.
We find the minima in horizontal (constant $\phi_{y^\prime}$) or vertical (constant $\phi_{x^\prime}$) slices.
The minimum for each vertical slice in Fig.~\ref{fig:radial_method1_extraction} is shown in red and is extracted from a profile similar to that shown in Fig.~\ref{fig:z_compensation}(a).
We then fit the minima with a straight line.
Finally, we repeat this procedure for $A_{y^\prime}(\phi_{x^\prime},\phi_{y^\prime})$ and find the intersection of the two fitted lines, achieving sub-noise precision in the radial micromotion minimum.
This example is given for Method 1, but the procedure is analogous for Methods 2 and 3.

\section{Precision metrics}\label{appendix:precision_metrics}
Here we explain how the values in Table~\ref{tab:precision_metrics} were obtained. For each method, we extract the uncertainty  $\Delta \phi_{i^\prime}$ in each compensation voltage required to nullify the stray field.
From the Method 1 data in Fig.~\ref{fig:method_123_2d}(a), for example, we extracted $\Delta \phi_{x^\prime} = \SI{9}{\volt}$ and $\Delta \phi_{y^\prime} = \SI{20}{\volt}$.

In the first column in Table~\ref{tab:precision_metrics}, the localization precision is $\Delta u_{0i^\prime} = \Delta(\mathbf{A}\cdot  \vec{\phi})_{i^\prime}$, where $\mathbf{A}$ is defined in Appendix~\ref{appendix:transformation}.
In the second column, the residual excess micromotion amplitude is $\frac{1}{2}u_{0i}q_i$ (see Eqn.~\eqref{eqn:eom_1d_solution}). As the value for $u_{0i}$, we use the localization precision $ \Delta u_{0i^\prime}$. As the value for $q_i$, we use $\frac{2\omega_i \sqrt{2}}{\Omega} $ since our radial Mathieu parameters satisfy $|a_i| \ll q_i^2/2$.
In the third column, the residual stray field uncertainty is given by $\eta_{i^\prime} \Delta {\phi}_{i^\prime}$, where $\eta_{i^\prime}$ is the field component at the trap center along the $i^\prime$ direction per unit ${\phi}_{i^\prime}$. From finite-element simulations of our trap, we obtain the values $(\eta_{x^\prime}, \eta_{y^\prime}, \eta_{z^\prime})= (0.5, 0.4, 6.6)~\si{\per \metre}$.

\section{Method 3 in different Paul traps}\label{appendix:diamond_trap}
\begin{figure}[hb]
	\includegraphics[width=1\linewidth]{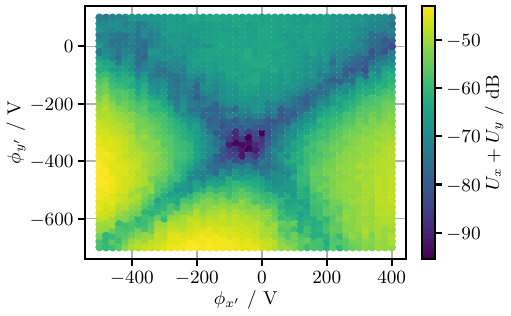}
	\caption{
		Data from applying Method 3 to find the micromotion minimum in the radial plane for a different Paul trap than that described in the main text.
		The setpoints $\phi_{x^\prime}$ and $\phi_{y^\prime}$ are shown on the horizontal and vertical axes, and the sum of the secular-motion amplitudes $U_{x}$ and $U_{y}$ during simultaneous tickling of the $x$ and $y$ modes are shown in the color bar.
	}
	\label{fig:diamond_trap_method3}
\end{figure}

Figure~\ref{fig:diamond_trap_method3} shows data from applying Method 3 to find the micromotion minimum in the radial plane for a different Paul trap than that described in the main text.
The particle was a cluster of several nanodiamonds of \SI{100}{\nano\meter} diameter (the mass was not determined), trapped in a linear blade-style Paul trap driven at $\Omega/2\pi=\SI{15}{\kilo\hertz}$, with a peak-to-peak amplitude of \SI{571}{\volt} and with \SI{810}{\volt} applied to both endcaps.
The modulation depth at $\Omega_{\text{tickle},i}=\omega_i$ was \SI{0.45}{\percent}.
After minimizing the micromotion, we estimated a residual excess micromotion amplitude of $\sim 1\si{\nano\meter}$.

In a third Paul trap in our laboratory, similar results were also obtained with a silica nanoparticle \SI{300}{\nano\meter} in diameter.

\end{document}